\documentclass[printer]{aa}
\usepackage[varg]{txfonts}
\usepackage{txfonts}
\usepackage{amsmath}

\usepackage[colorlinks]{hyperref}
\hypersetup{
    citecolor = {blue},
    linkcolor = {blue}
}

\begin{document} 
\title{
Comet 67P/Churyumov-Gerasimenko rotation changes derived from sublimation induced torques.
}
\titlerunning{67P/C-G rotation changes}

\author{T. Kramer \inst{1,2}
\and M. L\"auter \inst{1}
\and S. Hviid \inst{4}
\and L. Jorda \inst{3}
\and H.U. Keller \inst{4}
\and E. K\"uhrt \inst{4}
}

\institute{Zuse Institute Berlin, Supercomputing Division, Takustr.\ 7, 14195 Berlin, Germany
\and
Department of Physics, Harvard University, 17 Oxford St, Cambridge, MA 02138, USA
\and
Laboratoire d’Astrophysique de Marseille, UMR7326 CNRS/Université Aix-Marseille, 38 rue Frédéric Joliot-Curie, 13388 Marseille Cedex 13, France
\and
Institute of Planetary Research, Deutsches Zentrum für Luft- und Raumfahrt (DLR), Rutherfordstrasse 2, 12489 Berlin, Germany
}

\date{Received \today}

\abstract
   {
   The change of the rotation period and the orientation of the rotation axis of
   comet 67P/Churyumov-Gerasimenko (67P/C-G) is deducible from images taken by the scientific imaging instruments on-board the Rosetta mission with high precision.
   Non gravitational forces are a natural explanation for these data.
   }
   {We describe observed changes for the orientation of the rotation axis and the rotation period of 67P/C-G.
   For these changes we give an explanation based on a sublimation model with a best-fit for the surface active fraction (model~P).
   Torque effects of periodically changing gas emissions on the surface are considered.
   }
   {We solve the equation of state for the angular momentum in the inertial and the body-fixed frames and provide an analytic theory of the rotation changes in terms of Fourier coefficients, generally applicable to periodically forced rigid body dynamics.
   }
   {The torque induced changes of the rotation state constrain the physical properties of the surface, the sublimation rate and the local active fraction of the surface.}
   {We determine a distribution of the local surface active fraction in agreement with the rotation properties, period and orientation, of 67P/C-G.
   The torque movement confirms that the sublimation increases faster than the insolation towards perihelion.
   The derived relatively uniform activity pattern is discussed in terms of related surface features.
   }


   \maketitle
%
%________________________________________________________________

\section{Introduction}

Rosetta's instruments have probed the gas and dust environment during almost the entire apparition of 67P/Churyumov-Gerasimenko (67P/C-G) from 2014-2016.
The observed changes of the rotation period and axis orientation provide an independent measure of the sublimation activity of the nucleus by the induced torque.
The rotation period of 67P/C-G shortened by 21~min, the same number has been reported for the previous apparition by \cite{Mottola2014}.
\cite{Keller2015} proposed a sublimation driven model and explain the changing rotation period
in terms of a homogeneous ice model of the entire surface (\cite{Keller2015a}) without studying the axis orientation.
For many comets, changes in the rotation axis and of the orbital elements due to activity are observed and predicted, see \cite{Whipple1950,Jewitt1997,Samarasinha2004,Mueller2018}.
Before the shape of 67P/C-G was known in detail, \cite{Gutierrez2003} explored several scenarios for rotation-axis changes of small, irregularly shaped comets.
Typical changes of the rotation axis caused by sublimation forces range from $0.1$ to several tens of degrees.
For 67P/C-G the observed change is on the lower end of this range (0.5$^\circ$).

For the model~A proposed by \cite{Keller2015}, our analysis predicts a five times larger change of the direction of the rotation axis, also along a different direction compared to the observations.
In addition, model~A predicts a slower than observed increase of the total gas production with decreasing heliocentric distance, while the analysis of the coma by \cite{Hansen2016,Kramer2017,Lauter2018} shows a faster increase of the activity based on the pressure sensors.
To overcome these discrepancies, we establish a formalism to match models and observations in terms of a Fourier analysis of the gas induced torque and derive a possible ice distribution on the surface which explains the rotation-period changes, the movement of the {\color{black}direction of the} angular momentum, and the increase of activity with heliocentric distance (model~P).

\section{Forced rigid body dynamics}

We review the response of the rotation state of the nucleus to sublimation and other processes which alter the rotation and motion, see \cite{Thomson1986}.
The comet is viewed as a moving and rotating rigid body.
The cometary nucleus is defined within the three-dimensional body frame by a prescribed
\emph{body-frame density} $\rho_{\rm bf}(\vec{x},t)$ for each $\vec{x}$ in the body-frame at time $t$ to accommodate slow changes within the internal mass distribution.
The body frame is linked to the inertial frame by the coordinate transformation at time $t$
\begin{equation}\label{eq:trafo}
\vec{x}'(\vec{x},t) = \vec{r}(t) + \tens{R}(t) \vec{x},
\end{equation}
for each body-frame point $\vec{x}$.
Here, $\vec{r}(t)$ denotes the \emph{center of figure},
$\tens{R}(t)$ the orthogonal rotation matrix with the property
\begin{equation}\label{eq:rotmatrix}
\dot{\tens{R}}(t) \vec{x} = \vec{\omega}(t)\times \tens{R}(t)\vec{x},
\end{equation}
and $\vec{\omega}(t)$ the angular velocity.
The mapping $\vec{x}'(\vec{x},t)$ reflects the movement of the center of figure with time $t$, but leaves the shape geometry (defined in the body frame) unchanged.
The time-dependent density $\rho_{\rm bf}(\vec{x},t)$ allows one to incorporate slow changes in the density and porosity of the comet.
The density in the inertial frame $\rho(\vec{x}',t) = \rho_{\rm bf}(\vec{x},t)$
is obtained from the body-frame density considering Eq.~\eqref{eq:trafo} and carries along the time-dependence of the orbital and rotational movement.
The comet mass $M$ and the center of mass in the body frame $\bar{\vec{x}}$ are in general time-dependent
\begin{eqnarray}
M(t) &=& \int {\rm d} \vec{x}'\; \rho(\vec{x}',t)
= \int {\rm d} \vec{x}\; \rho_{\rm bf}(\vec{x},t),\\
\bar{\vec{x}}(t) &=& \frac{1}{M}\int {\rm d} \vec{x}\; \vec{x} \,\rho_{\rm bf}(\vec{x},t).
\end{eqnarray}
For the center of mass in the inertial frame the relation holds
\begin{equation}
\overline{\vec{x}'}(t) = \frac{1}{M} \int {\rm d} \vec{x}'\; \vec{x}' \,\rho(\vec{x}',t)
= \vec{r} + R \bar{\vec{x}}.
\end{equation}
The time derivative of Eq.~\eqref{eq:trafo} yields the
inertial-frame velocity $\vec{V}$ of a fixed body-point
\begin{equation}
\vec{V}(\vec{x},t) = \partial_t \vec{x}' = \dot{\vec{r}} + \vec{\omega} \times \tens{R}\vec{x}.
\end{equation}
To obtain the linear momentum $\vec{P}$ and angular momentum $\vec{L}$ of the whole nucleus in the inertial system we integrate
\begin{eqnarray}%\label{eq:momentum}
\vec{P}(t) & =& \int {\rm d}\vec{x}' \rho(\vec{x}',t) \, \vec{V}(\vec{x}',t)
= M\dot{\vec{r}} + M \vec{\omega} \times \tens{R} \bar{\vec{x}}, \label{eq:momentum1} \\
\vec{L}(t) & = & \int {\rm d}\vec{x}' \rho(\vec{x}',t) \,
(\vec{x}'-\vec{r})\times \vec{V}(\vec{x}',t) 
=  \tens{I} \vec{\omega} - M \dot{\vec{r}} \times \tens{R} \bar{\vec{x}},\label{eq:momentum2}
\end{eqnarray}
with the tensor of inertia $\tens{I}(t) = \tens{R} \tens{I}_{\rm bf}(t) \tens{R}^{-1}$
in the inertial frame with respect to the center of figure $\vec{r}$
and the tensor of inertia $\tens{I}_{\rm bf}(t)$ with respect to the body-frame center $0$.
For the case of a time-dependent body density, $\tens{I}_{\rm bf}(t)$ needs to be computed for the body-frame density at time $t$.
The momentum changes are generated by the sum of non-gravitational and gravitational forces $\vec{F}_{\rm NG} + \vec{F}_{\rm G} = \dot{\vec{P}}$ and torques $\vec{T}_{\rm NG} + \vec{T}_{\rm G}= \dot{\vec{L}}$.
Gas sublimation at point $\vec{x}$ on the surface leads to a mass loss $\dot{m}$ and generates
the non-gravitational components
\begin{eqnarray}\label{eq:force}
\vec{F}_{\rm NG}(t) & =& \int {\rm d}\sigma'\, \dot{m}(\vec{x}',t) \vec{v}_{\rm is}(\vec{x}',t)\nonumber \\
&=& \tens{R} \int {\rm d}\sigma\, \dot{m}(\vec{x},t) \vec{v}_{\rm bf}(\vec{x},t) , \label{eq:force1}
 \\
\vec{T}_{NG}(t) & =& \int {\rm d}\sigma'\, \dot{m}(\vec{x}',t) (\vec{x}'-\vec{r}) \times \vec{v}_{\rm is}(\vec{x}',t) \nonumber \\
&=& \tens{R} \int {\rm d}\sigma\,  \dot{m}(\vec{x},t) \vec{x} \times \vec{v}_{\rm bf}(\vec{x},t).
\label{eq:force2}
\end{eqnarray}
The gas velocity in the inertial frame $\vec{v}_{\rm is} = u_{\rm gas} \hat{\vec{n}}_0 + \vec{\omega}\times (\vec{x}'-\vec{r})$ consists of two components, one into normal direction $\hat{\vec{n}}_0(\vec{x}',t)$ on the nucleus' surface in the inertial frame and one due to the body rotation, $u_{\rm gas}$ denotes the thermal gas velocity from Eq.~(\ref{eq:vth}).
The gas velocity in the body frame is given by
$\vec{v}_{\rm bf}
= \tens{R}^{-1} \vec{v}_{\rm is} = u_{\rm gas}\hat{\vec{n}} + (\tens{R}^{-1}\vec{\omega}) \times \vec{x}$
with the outward surface normal $\hat{\vec{n}}(\vec{x},t)$ at surface location $\vec{x}$
within the body frame.
According to \cite{Jorda2002}, Eq.~(11), the mass production $\dot{m}$ for a mixture of
{\color{black}
gas species reads
\begin{equation}\label{eq:massloss}
\dot{m}(\vec{x},t) =  \sum_{\rm gas} f_{\rm gas}(\vec{x}) Z_{\rm gas}(\vec{x},t),
\end{equation}
with the surface active fraction $f_{\rm gas}$ and the sublimation rate $Z_{\rm gas}$.
The mass loss $\dot m$ changes the shape, reduces the total mass, and affects the tensor of inertia of the nucleus.
The integrated mass loss of comet 67P/C-G during the 2015 apparition is estimated to be about 1/1000 of the total mass (see \cite{Godard2015} and \cite{Godard2017}) and therefore we neglect both effects, we assume time-independent mass and tensor of inertia.
}
{\color{black}
The gravitational components for force and torque yield
\begin{eqnarray}
\vec{F}_{\rm G}(t) & =& \int {\rm d}\vec{x}'\, \rho(\vec{x}',t) \vec{a}(\vec{x}',t)
= M \vec{a}(\overline{\vec{x}'},t), \label{eq:gforce1}
 \\
\vec{T}_{\rm G}(t) & =& \int {\rm d}\vec{x}'\, \rho(\vec{x}',t) (\vec{x}'-\vec{r}) \times
\vec{a}(\vec{x}',t) \nonumber\\
&=& M (\tens{R} \bar{\vec{x}})\times \vec{a}(\overline{\vec{x}'},t) \label{eq:gforce2}
\end{eqnarray}
with the gravitational acceleration $\vec{a}$ due to other solar system bodies.
For both volume integrals in Eqs.~\eqref{eq:gforce1}, \eqref{eq:gforce2}, $\vec{a}(\vec{x}')$ is assumed to be constant over the cometary body, which implies that e.g.\ tidal forces are neglected.
}

Eqs.~\eqref{eq:rotmatrix}, \eqref{eq:momentum1}, \eqref{eq:momentum2},
\eqref{eq:force1}, \eqref{eq:force2}, \eqref{eq:gforce1}, \eqref{eq:gforce2}
result in a system of coupled algebraic and differential equations
\begin{eqnarray}\label{eq:dynamics}
\dot{\vec{P}} = \vec{F}_{\rm NG} + M\vec{a},\quad
\dot{\tens{R}}() = \vec{\omega} \times \tens{R}(),\quad
\dot{\vec{L}} = \vec{T}_{\rm NG} + M \tens{R}\bar{\vec{x}}\times \vec{a},\\
M \dot{\vec{r}} + M\vec{\omega}\times \tens{R}\bar{\vec{x}} = \vec{P},\quad
\tens{I}\vec{\omega} - M\dot{\vec{r}} \times \tens{R}\bar{\vec{x}} = \vec{L}
\end{eqnarray}
for the state variables $\vec{r}(t)$, $\vec{P}(t)$, $\tens{R}(t)$, $\vec{L}(t)$.
Eqs.~\eqref{eq:momentum1}, \eqref{eq:momentum2} couple linear and rotational momenta through the center of mass in the body frame $\bar{\vec{x}}(t)$.
If the density distribution $\rho_{\rm bf}$ satisfies $\bar{\vec{x}}=0$,
$\vec{r} = \overline{\vec{x}'}$ becomes the center of mass in the inertial frame and  Eqs.~\eqref{eq:dynamics} decouple into two blocks.
The first block
\begin{equation}\label{eq:lindyn}
\dot{\vec{P}} = \vec{F}_{\rm NG} + M \vec{a},\quad
M \dot{\vec{r}} = \vec{P}
\end{equation}
describes the translational movement for the state variables $\vec{r}(t)$, $\vec{P}(t)$,
and the second one
\begin{equation}\label{eq:rotdyn}
\dot{\tens{R}}() = \vec{\omega} \times \tens{R}(),\quad
\dot{\vec{L}} = \vec{T}_{\rm NG},\quad
\tens{I}\vec{\omega} = \vec{L}
\end{equation}
the rotational dynamics for $\tens{R}(t)$ and $\vec{L}(t)$.
The model for the changing rotation period by \cite{Keller2015} is contained as special case in Eq.~\eqref{eq:rotdyn}.
For that let us denote the eigenvector $\vec{e}$ of $I_{\rm bf}$ in the body frame with the largest moment of inertia
$I_{\rm z}$  and assume $\vec{L}$ initially aligned with 
$\vec{e}' = \tens{R}\vec{e}$, that is
$\vec{L}(0) = L_{\rm z} \vec{e}'(0)$.
Then $\vec{\omega}(0) = L_{\rm z}/I_{\rm z}\vec{e}'(0)$ and consequently
$\dot{\tens{R}} \vec{e} = 0$.
Thus, $\vec{e}' = \tens{R}\vec{e}$ is constant in time and the angular velocity changes with $\dot{\vec{\omega}}\cdot \vec{e}' = \vec{T}_{\rm NG}\cdot \vec{e}' / I_{\rm z}$.

For given initial conditions of all state variables, the system of Eqs.~\eqref{eq:dynamics}, \eqref{eq:lindyn}, and \eqref{eq:rotdyn} can be solved numerically.
For accuracy, we use the LSODE package provided within Mathematica/FORTRAN (\cite{Radhakrishnan1993}).
Following \cite{Shoemake1985}, the matrix-matrix operation
$\tens{R}\rightarrow\vec{\omega} \times \tens{R}$
is replaced by a quaternion multiplication for improved stability.
There is a one-to-one mapping between $\tens{R}$ and a quaternion $\vec{q}$
such that the matrix matrix operation is substituted by
$\vec{q}\rightarrow \vec{q}\cdot (0,\vec{\omega})/2$.

\section{Rotation state of 67P/C-G}

After the arrival of Rosetta at 67P/C-G in 2014, observations of the rotation by \cite{Preusker2015} and \cite{Godard2017} show 67P/C-G in an excited state of rotation, albeit with the rotation axis close to the axis $\vec{e}'$ with the largest moment of inertia (rotation state with minimum energy).
This points to an alignment of both axes which is also compatible with observations (\cite{Jorda2016}).
{\color{black}The total mass is estimated to be $10^{13}$~kg by \cite{Godard2015}.
The assumption of a strictly homogeneous density leads to a tensor of inertia 
\begin{eqnarray}\label{eq:inertiaHOM}
\tens{I}_{\rm hom}=\left(\!\!\!
\begin{array}{ccc}
 9.55529\times 10^{18} & \!\!\! 1.73767\times 10^{16} &  \!\!\!2.24462\times 10^{17} \\
 1.73767\times 10^{16} & \!\!\! 1.76369\times 10^{19} & \!\!\!  -7.45958\times 10^{16} \\
 2.24462\times 10^{17} & \!\!\! -7.45958\times 10^{16} & \!\!\! 1.89825\times 10^{19} \\
\end{array}
\!\!\!\right) \\\nonumber \text{kg m$^2$}.
\end{eqnarray}
with respect to the center of mass.
This tensor is incompatible with the observations,} since then the axis $\vec{e}'$ would be tilted by $2.9^\circ$ with respect to the rotation axis (\cite{Preusker2017}), and in addition an offset of the center of mass
$\overline{\vec{x}'}$ with respect to the center of figure $\vec{r}$ would exist (\cite{Jorda2016}).
{\color{black}Inhomogeneities in the density have also been reported by
\cite {Brouet2016} and \cite{Knapmeyer2018} from the CONSERT
and SESAME/MUPUS Rosetta data.}
In the following we assume a {\color{black}time-independent}, non-homogeneous density distribution which aligns the rotation axis with the $\vec{e}'$ under the constraint that the total mass is kept fixed at $M=10^{13}$~kg and $\overline{\vec{x}'} = \vec{r}$.
For definiteness we give a possible mass distribution with resulting tensor of inertia
\begin{eqnarray}\label{eq:inertia}
\tens{I}_{\rm bf}=
\left(\!\!\!
\begin{array}{ccc}
 9.3408457\times 10^{18} & \!\!\! 5.6695663\times 10^{16} & \!\!\! 0 \\
 5.6695663\times 10^{16} & \!\!\! 1.6562414\times 10^{19} & \!\!\! 0 \\
 0 & 0 & \!\!\!\!\!\!\!\!\!\!\!\!\!\!\! 1.8192083\times 10^{19} 
\end{array}
\!\!\!\right) \\\nonumber \text{kg m$^2$}.
\end{eqnarray}
This solution equals putting $(1/11) M$ on a thin ring centered at $\{-159.8, 275.5, -220.5\}$~m in the plane $-0.4065601x+ 0.01699493y + 0.9134659z = -131.7705$~m with radius $1$~km, while distributing $(10/11) M$ homogeneously throughout the entire nucleus.
The solution is not unique and leads to an increased density in the big lobe, in agreement with \cite{Jorda2016}.
{\color{black}
The small off-diagonal entries in Eq.~(\ref{eq:inertia}) align the shape file with the body frame by a rotation of $0.4^\circ$.
The changes of the rotation state of the comet is determined by the relative change of the angular momentum with respect to the initial state.
Changing the total mass does not affect the resulting dynamics, if the average surface active fraction is changed in the same proportion.
}

\subsection{Sublimation}\label{subsec:sublimation}

\begin{figure}
\begin{center}
\includegraphics[width=0.995\linewidth]{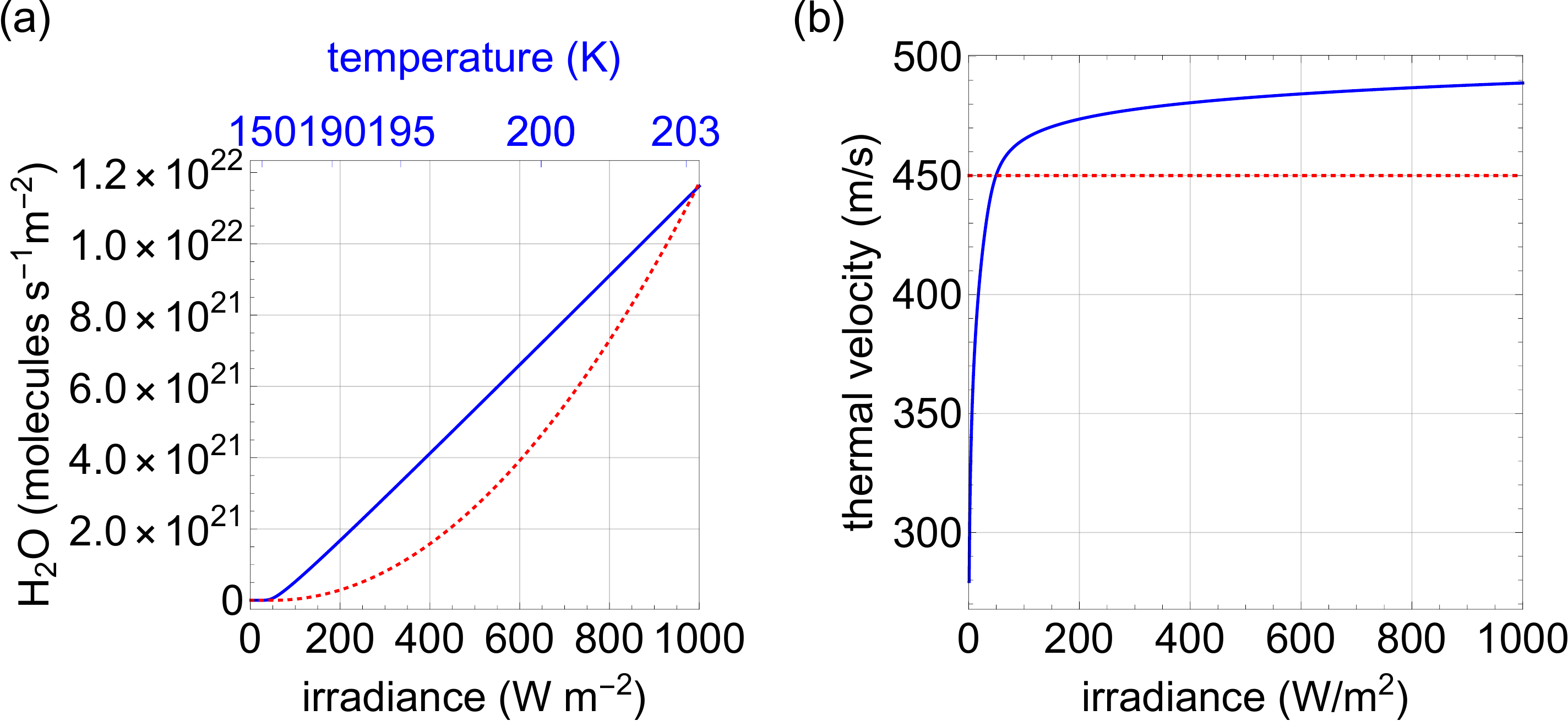}
\end{center}
\caption{Radiation driven sublimation (a) rate and (b) velocity as function of received irradiation.
Solid blue lines: model~A (surface active fraction $=1$),
dashed red lines: effective sublimation curve with enhanced radiation response near perihelion.
The effective sublimation curve leads to agreement with observations (see Figs.~\ref{fig:resultPole},\ref{fig:resultPeriodProduction}).
}
\label{fig:subcurve}
\end{figure}

Changes in the rotation state are primarily due to the torque induced by sublimation of ice.
The total production of water and $CO_2$ has been estimated by \cite{Hansen2016,Lauter2018} from ROSINA COPS and DFMS data to be about $6.2\pm 2\times 10^9$~kg, corresponding to about $1/1600$ of the total mass of 67P/C-G ($M=10^{13}$~kg).
The water production shows a steep increase with heliocentric distance $r_h^{\alpha}$ around Southern solstice, with exponents $\alpha$ ranging from $-6.5$ up to $-7$.
The total gas production from the radiation driven sublimation model~A by \cite{Keller2015} yields smaller exponents $\alpha=-2.8$.
{\color{black}
Our model for the rotation state only considers water emission from the surface for driving the torque evolution.
The $CO_2$ activity liberates decimeter sized chunks (\cite{Keller2017}) that contain additional water which is seen as  production but does not contribute to the torque and does not have the same diurnal signature as the surface.
The $CO_2$ contribution (about 1/7 of the water mass estimated from ROSINA/DFMS by \cite{Lauter2018}) is not considered separately since the $CO_2$ sources around perihelion coincide with the water regions (\cite{Fougere2016a}, \cite{Lauter2018}) and drive the torque in a similar direction as water.
In addition the diurnal variation of $CO_2$ is less pronounced than for water (see \cite{Filacchione2016}) and thus has less influence on the periodic torque components. 
At heliocentric distances larger than $3$~a.u., and in particular on the outbound orbit, $CO_2$ becomes the dominant species (\cite{Lauter2018}) and does not follow the subsolar illumination.
At these distances the rotation period of the comet has settled and these times are outside the scope of the present analysis.}
The rotation axis of 67P/C-G shows largest movements around $\pm 100$~days from perihelion, in agreement with the larger $|\alpha|$ exponents derived from the gas instruments.
\cite{Keller2015} (model~A) consider the gas production based on a shape model.
On each surface element with a given Bond bolometric albedo $A$ and solar irradiance $f_\sun=S_\sun/r_h^2$ at heliocentric distance $r_h$~[a.u], solar constant $S_\sun=1361$~W\,m$^{-2}$,
{\color{black}
the energy balance
\begin{equation}\label{eq:Zenergy}
(1 - A) f_\sun =\epsilon \sigma T^4 + Z(T) L_{\rm ice}
\end{equation}
is solved for the sublimation rate $Z$, given by the Hertz-Knudsen relation
}
\begin{equation}
Z_{\rm Hertz-Knudsen}(T) =  2 P(T)/(\pi v_{\rm th}).
\label{eq:ZHK}
\end{equation}
The parameters are taken from \cite{Keller2015}, with emissivity $\epsilon=0.9$, latent heat of sublimation for water ice $L_{\rm ice}=2.6\times 10^6$~J\,kg$^{-1}$ (assumed to be constant), water vapor pressure $P(T)=3.56\times10^{12} \,{\rm e}^{-6141.667/T}$~[kg\,m$^{-1}$\,s$^{-2}$], and thermal velocity of water molecules with molar mass $\mu_{H_2O}$ 
\begin{equation}\label{eq:vth}
v_{\rm th}(T)=\sqrt{8 R T/(\pi\mu_{H_2O})}
\end{equation}
The gas constant is denoted by $R$, the Stefan-Boltzmann constant by $\sigma$.
The solution to Eqs.~(\ref{eq:Zenergy},\ref{eq:ZHK}) in terms of the sublimation rate $\dot{N}\equiv Z_{\rm Hertz-Knudsen}/m_{H_2O}$ [s$^{-1}$\,m$^{-2}$], mass of water molecule $m_{H_2O}$~[kg],
is shown in Fig.~\ref{fig:subcurve} for $A=0.01$ (\cite{Keller2015}, model~A).
The observed changes by \cite{Hansen2016,Kramer2017,Lauter2018} point to a faster increase of the total production with decreasing perihelion distance.
{\color{black}
To account for the observation requires to assume a sublimation rate which increases faster than linearly with illumination, as exemplified by the dashed red line in Fig.~\ref{fig:subcurve}.
}
A possible physical mechanism behind the increase of the sublimation could be a decrease of the dust layer when the comet approaches the sun, leading to a steeper slope (model~C to model~A transition by \cite{Keller2015}).
{\color{black}The required adjustment of the sublimation rate with heliocentric distance has also been suggested by \cite{Marsden1973} and is used in the DSMC coma models for 67P/C-G by (\cite{Fougere2016}, Eq.~(1)).}

\subsection{Observed rotation axis changes}

\begin{figure*}
\begin{center}
\includegraphics[width=0.9\linewidth]{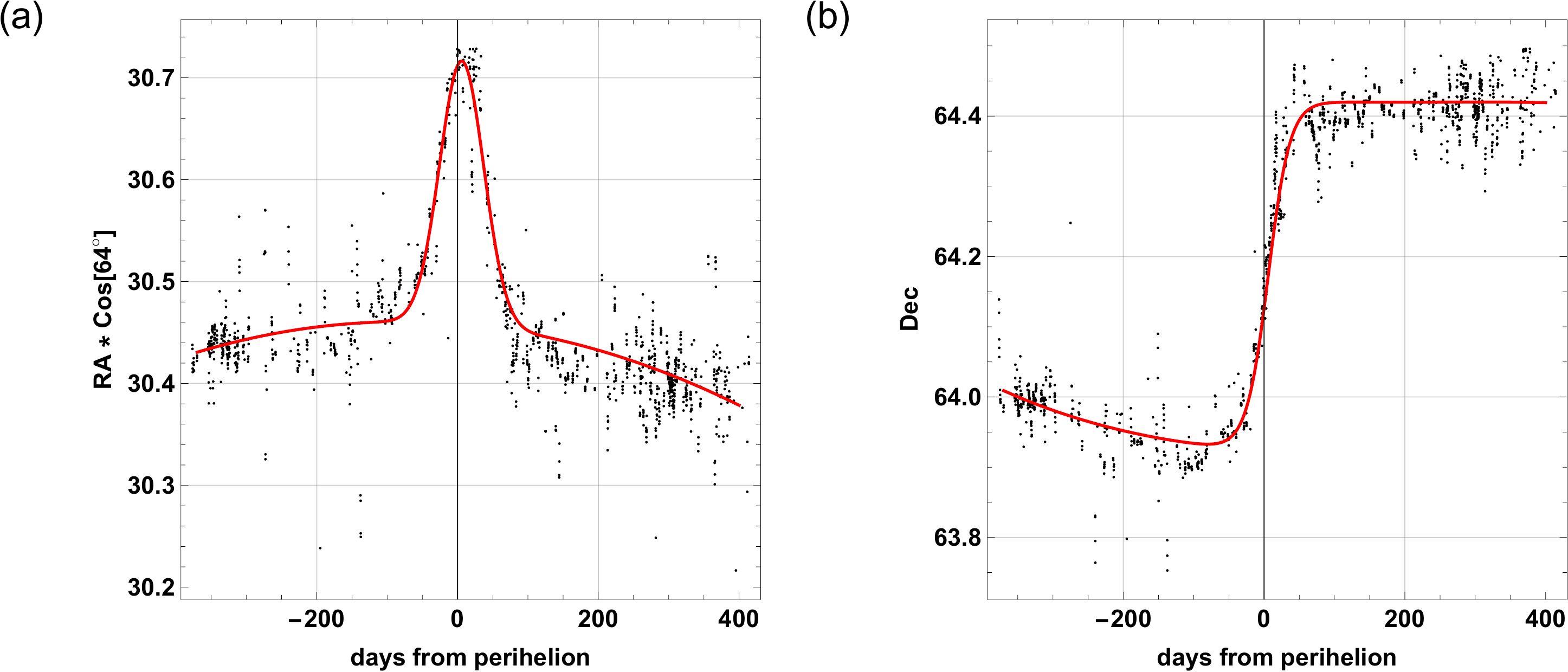}
\end{center}
\caption{Orientation changes of the rotation axis.
Black dots denote the reconstructed right ascension {\color{black}multiplied with $\cos 64^\circ$} and declination, the solid line a smooth fit to the data.
}
\label{fig:radec}
\end{figure*}

The (RA,Dec) orientation of the angular velocity $\vec{\omega}$ of the comet nucleus is derived using the set of about 25000 control points defined as the center of the maplets created in the stereophotoclinometry (SPC) method of \cite{Gaskell2008} applied to comet 67P/C-G by \cite{Jorda2016}. 
The coordinates of the stereo control points measured on sequences of Rosetta/OSIRIS images (\cite{Keller2007}) combined with star tracker pointing measurements are used to determine the direction of the angular velocity vector in the Equatorial J2000 (EME2000) reference frame during the Rosetta mission, see Fig.~\ref{fig:radec}. 
The fluctuations in the resulting data set are caused by a possible nutation combined with the uncertainties in the determination of the direction of the $\vec{\omega}$ vector.
They led us to strongly smooth the data as illustrated by the solid line in Fig.~\ref{fig:radec}, which represents the time-averaged motion of the rotation axis by fitting the data to a Gaussian function for the RA and to a hyperbolic tangent function for the Dec, in addition to a third order polynomial.
A similar data set has been retrieved by \cite{Godard2017}.

\section{Fourier theory of torque induced motion}

The shape model from \cite{Preusker2017} is re-meshed using the Aproximated Centroidal Voronoi Diagrams (ACVD) tool by \cite{Valette2008} into $N_{\rm faces}=3996$ triangular elements with area $A_i$ and surface normal $\hat{\vec{n}}_i$, chosen to be approximately of equal area.
The results are robust against variations of the shape model as long as $N_{\rm faces}\gg 100$.
The torque is evaluated according to Eqs.~(\ref{eq:force2}), (\ref{eq:massloss}).
The sublimation rate $Z_{\rm gas}$ on each facet is evaluated for a given subsolar latitude $\phi_s$ and heliocentric distance $r_h$ during one rotation period, starting with subsolar longitude $\lambda_s=0$.
{\color{black}
We compute the total torque arising from the water sublimation curves, either the Hertz-Knudsen rate from Eq.~(\ref{eq:ZHK}) or alternatively the effective sublimation curve in Fig.~\ref{fig:subcurve}:}
\begin{equation}\label{eq:Tbf}
\vec{T}_{{\rm bf}}(\lambda_s,\phi_s,r_h)=
\sum_i^{N_{\rm faces}} f_i A_i Z_{{H_2O},i}\,v_{{\rm th},i}\,\vec{r}_i\times \hat{\vec{n}}_i .
\end{equation}
The contributions of shadowed (including self-shadowed) faces are set to zero and all the surface active fraction are first set to the same value.
In the next iteration, the surface active fraction is considered to be a spatial fit parameter (but fixed in time) to be determined from the observed axis changes.
\begin{figure}
\begin{center}
\includegraphics[width=0.9\linewidth]{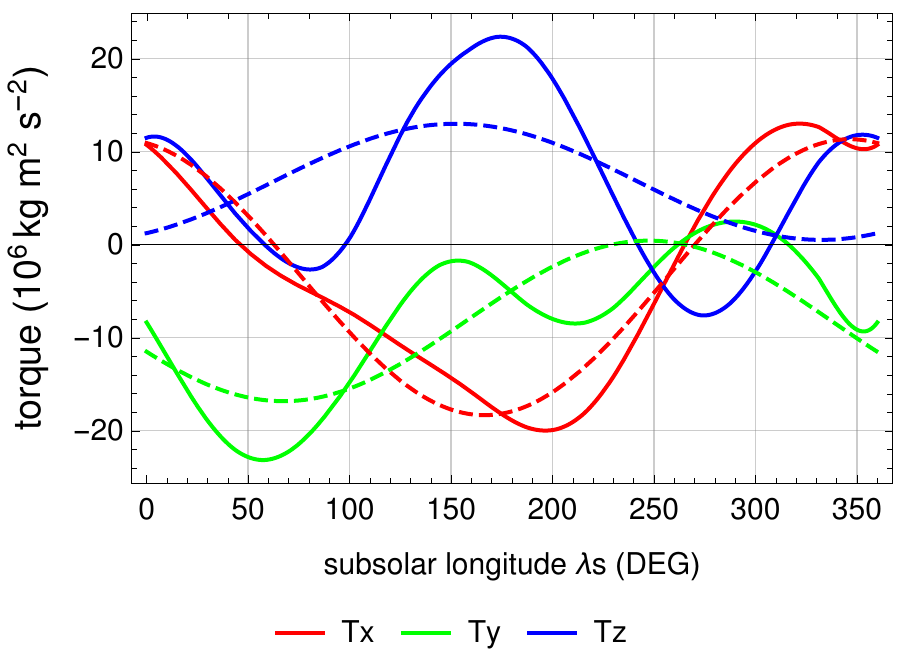}
\end{center}
\caption{{\color{black}Calculated sublimation induced torque at perihelion for one rotation period at perihelion as function of subsolar longitude from model~P using the effective sublimation curve from Fig.~\ref{fig:subcurve} with the best fit solution from Fig.~\ref{fig:awf}.
The dashed lines show the Fourier representation (constant, $\cos$, $\sin$ terms) of the corresponding solid lines to parametrize the diurnal cycle.}
}\label{fig:diurnalTorque}
\end{figure}
The heliocentric orbit (cartesian coordinates $r_h(t)$) is taken from the NASA Horizon system (Earth Mean Equator and Equinox of Reference Epoch J2000).
The initial orientation of the rotation axis in the equatorial frame is set to be in the lowest energy state (rotation axis and angular momentum aligned, pointing to RA $\alpha=69.427^\circ$, Dec $\delta=64.000^\circ$ at $t=-350$~days).
For 67P/C-G the observed axis changes are small and we tabulate the subsolar latitude $\phi_s$ and heliocentric distance $|r_h|$ in $N_{\rm intervals}=81$ 10-days~intervals and store the body-frame torque as function of $\lambda_s$.
A typical diurnal torque evolution at perihelion is shown in Fig.~\ref{fig:diurnalTorque}.
To gain more physical insight into the dynamics underlying the axis changes, we expand the torque components in a Fourier cosine/sine series.
The periodic argument $\alpha=[-\pi:\pi[$ of the Fourier series is not time, but the subsolar longitude $\lambda_s=\alpha+\pi$ to accommodate changes in the rotation period.
In the $i$th time interval we extract the first $N=2 m+1$ Fourier coefficients $\vec{C}_n^{(i)}=\{C^{(i)}_{n,x},C^{(i)}_{n,y},C^{(i)}_{n,z}\}$,
which yield the Fourier series representation of the torque
\begin{equation}
\vec{T}_{F,{\rm bf}}^{(i)}(\alpha)=\vec{C}_{0}^{(i)}+\sum_{n=1}^m \vec{C}_{n}^{(i)}\sin (n\alpha)+\sum_{n=1}^m \vec{C}_{m+n}^{(i)}\cos (n\alpha)\\
\end{equation}
The final parametrization of the body torque as function of subsolar longitude along the entire orbit is given by the time evolution of the Fourier coefficients
\begin{equation}\label{eq:FCint}
\vec{C}^{\rm int}_n(t)=\text{interpolation}(\vec{C}^{(i)}_n,\ldots,\vec{C}_n^{(N_{\rm intervals})}),
\end{equation}
and rewriting
\begin{eqnarray}\nonumber
\vec{T}_{F,bf}(t,\lambda_s)
&=&\vec{C}_{0}^{\rm int}(t)
  +\sum_{n=1}^m \vec{C}_{n}^{\rm int}(t)  \sin n(\lambda_s-\pi)\\
&&+\sum_{n=1}^m \vec{C}_{m+n}^{\rm int}(t)\cos n(\lambda_s-\pi).\label{eq:FBF}
\end{eqnarray}
The slowly changing subsolar latitude and heliocentric distance are implicitly contained through the time $t$ argument.
The time evolution of the coefficients is shown in Fig.~\ref{fig:fcallA} for $N=3$.
The first row in Fig.~\ref{fig:fcallA} displays the non-diurnal torque coefficients.
Since the rotation axis is aligned with the $z$ body-axis, the torque component $C^{\rm int}_{0,z}(t)$ directly affects the rotation period.
The orientation changes of the rotation axis are caused by the diurnal $C^{\rm int}_{1,x}, C^{\rm int}_{2,x}, C^{\rm int}_{1,y}, C^{\rm int}_{2,y}$ components (Fig.~\ref{fig:diurnalTorque}).
The diurnal components are not linearly independent as discussed in Sect.~\ref{sec:Dynamics}.

\begin{figure*}
\begin{center}
\includegraphics[width=0.8\linewidth]{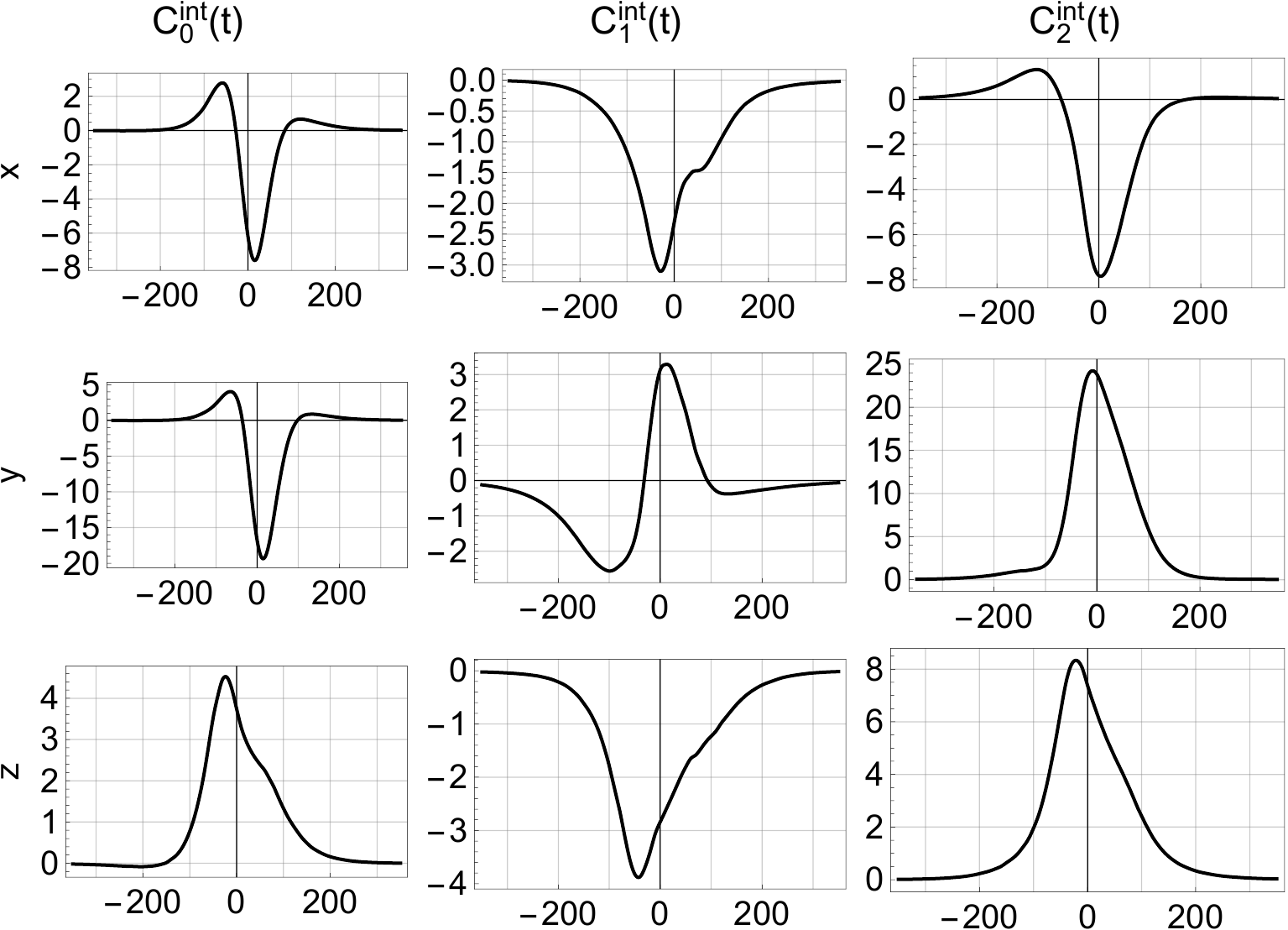}
\end{center}
\caption{
{\color{black}Torque evolution of a uniform active surface model with the effective sublimation curve from Fig.~\ref{fig:subcurve}}. 
The time evolution of the first 3 Fourier coefficients $\vec{C}_0^{\rm int}(t)$, $\vec{C}_1^{\rm int}(t)$, $\vec{C}_2^{\rm int}(t)$ is shown for $(-300:300)$~days from perihelion for each Cartesian component of the body frame torque, Eq.~(\ref{eq:FCint}).
Units of $10^6$~kg m$^2$ s$^{-2}$.
}
\label{fig:fcallA}
\end{figure*}

\begin{figure*}
\begin{center}
\includegraphics[width=0.8\linewidth]{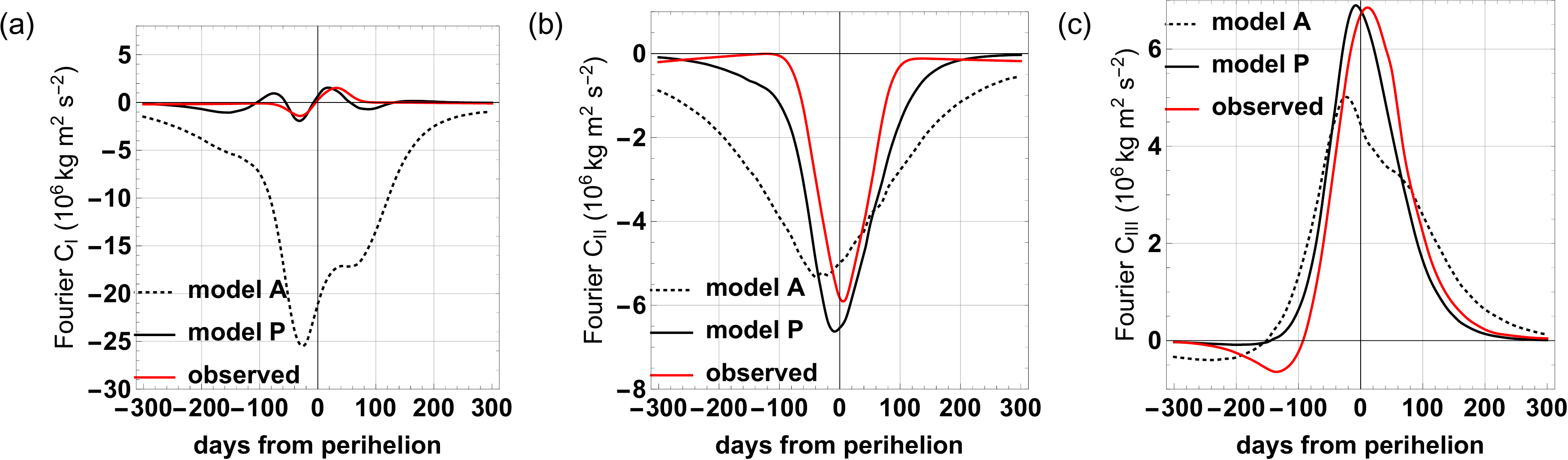}
\end{center}
\caption{Time evolution of the three physical relevant combinations of Fourier coefficients (Eq.~\ref{eq:FCrobust}) for $(-300:300)$~days from perihelion, units of $10^6$~kg m$^2$ s$^{-2}$.
(a) model~A, global uniform active surface $1/12$, 
(b) model~P (patches with effective sublimation curve),
(c) observed Fourier coefficients inferred from the rotation-axis movement and the tensor of inertia
}\label{fig:robustFC}
\end{figure*}

\subsection{Computation of the torque in the inertial system}\label{sec:Dynamics}

\begin{figure*}
\begin{center}
\includegraphics[width=0.9\linewidth]{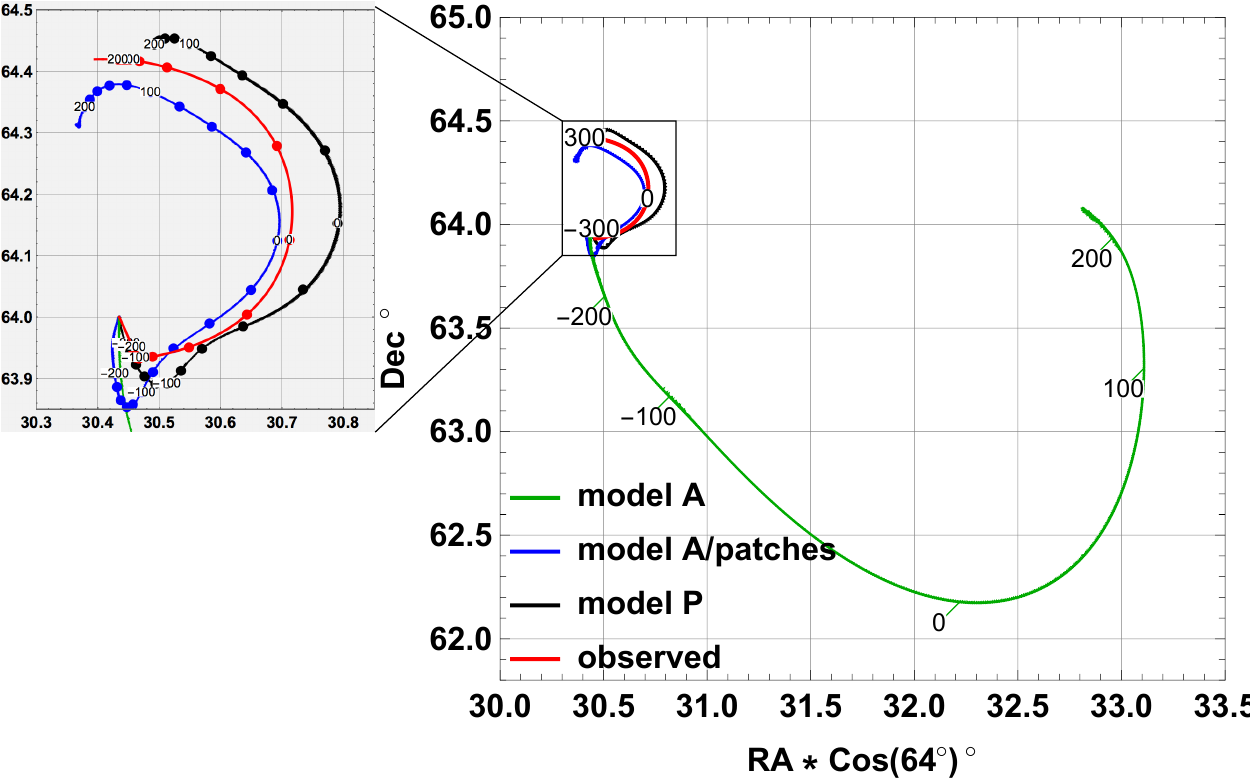}
\end{center}
\caption{Rotation-axis movement.
Red line: observation from Fig.~\ref{fig:radec}, 
other lines represent different sublimation models:
model~A with globally constant surface active fraction ($1/12$),
model~A/patches with best fit adjustment of patches,
model~P with effective sublimation curve and best fit adjustment of patches.
The grey inset shows a magnification of the curves.
}\label{fig:resultPole}
\end{figure*}

\begin{figure*}
\begin{center}
\includegraphics[width=0.9\linewidth]{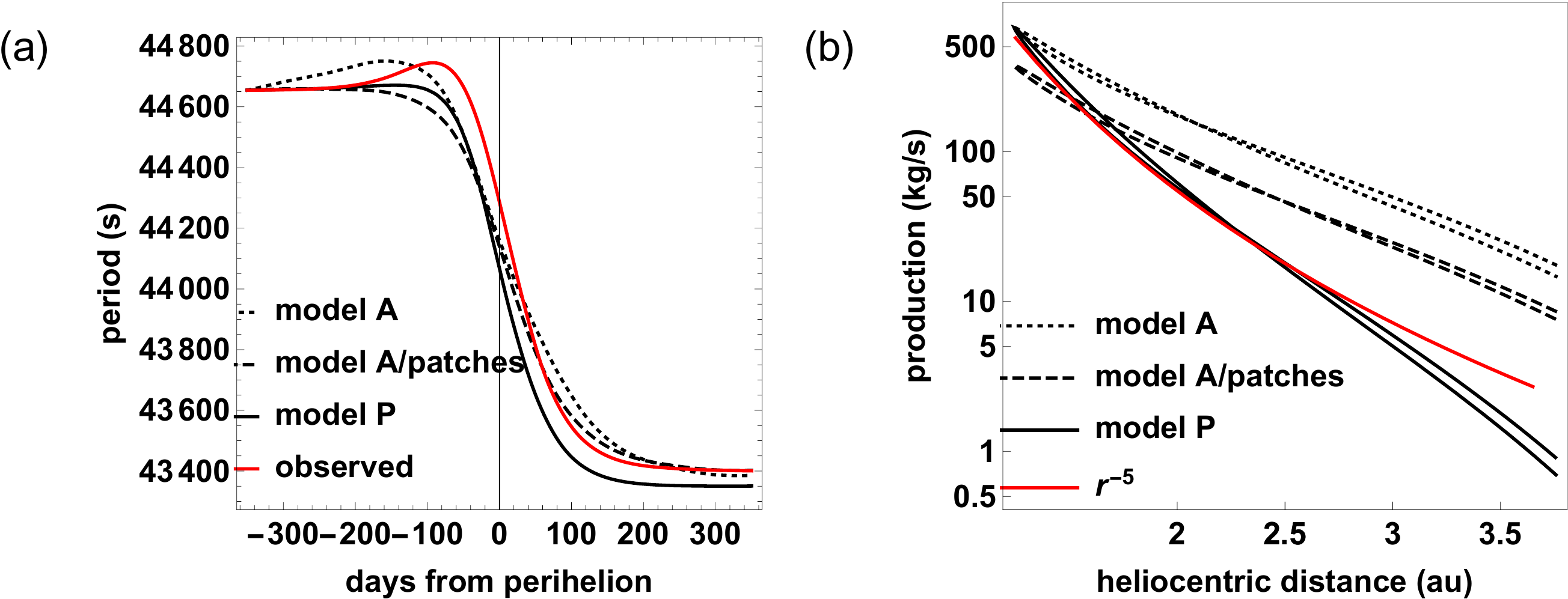}\\
\end{center}
\caption{Rotation period and total water production.
Red line: observation, black lines: different sublimation models:
model~A with globally constant surface active fraction ($1/12$),
model~A/patches with best fit adjustment of patches,
model~P with effective sublimation curve and best fit adjustment of patches.
In all cases, the rotation period is in reasonable agreement with observations.
The total water production rate drops for the model~A and model~A/patches scenarios with $r_h^{-2.8}$ while observations indicate $r_h^{-5}$.
}\label{fig:resultPeriodProduction}
\end{figure*}

First, we neglect the changes of the orbital elements due to non-gravitational momentum (Eq.~\ref{eq:lindyn}) and take the orbital evolution $\vec{r}_h(t)$ as fixed.
For a given rotation matrix $\tens{R}(t)$ (body-frame to equatorial inertial-system) and position of the comet $\vec{r}_h(t)$, the sub-solar longitude  is given by
\begin{eqnarray}\label{eq:lambdas}
\{x(t),y(t),z(t)\}&=&\tens{R}^{-1}(t) (-\vec{r}_h(t))\\
\lambda_s(t)&=&\arctan(y(t)/x(t))
\end{eqnarray}
At initial time $t_S$ the rotation matrix $\tens{R}(t_S)$, which transfers the body-frame $z$-axis to the rotation axis $\hat{\vec{s}}=\{s_x,s_y,s_z\}$ in the equatorial inertial frame, is given by
\begin{equation}
\tens{R}(t_S)=
\begin{pmatrix}
 \frac{s_z s_x^2+s_y^2}{s_x^2+s_y^2} & \frac{s_x s_y (s_z-1)}{s_x^2+s_y^2} & s_x
   \\
 \frac{s_x s_y (s_z-1)}{s_x^2+s_y^2} & \frac{s_x^2+s_y^2 s_z}{s_x^2+s_y^2} & s_y
   \\
 -s_x & -s_y & s_z
\end{pmatrix}.
\end{equation}
Eq.~\eqref{eq:rotdyn} is then integrated with the initial angular velocity and momentum set to
\begin{eqnarray}
\vec{\omega}_{\rm bf}(t_S) & = &\{0,0,2\pi/T_{\rm rot}\}, \quad T_{\rm rot}=44650\text{~s}, \\
\vec{L}(t_S)& = & \tens{R}(t_S) \tens{I}_{\rm bf}
\vec{\omega}_{\rm bf}(t_S).
\end{eqnarray}
The Fourier decomposition provides additional insights into the axis changes.
An important parameter is the angle $\lambda_s^0$ around the $z$-axis to point the $x$-$z$-plane towards towards the sun (zero sub-solar latitude).
Neglecting the $0.5^\circ$ tilt-change of the rotation axis, $\lambda_s^0$ is given by
\begin{eqnarray}
\{x_s^0(t),y_s^0(t),z_s^0(t)\}&=&\tens{R}(t_S)^{-1} (-r_h(t))\\
\lambda_s^0(t)&=&\arctan(y_s^0/x_s^0).
\end{eqnarray}
We obtain a good approximation of the angular momentum change $\boldsymbol{\Delta} \tilde{\vec{L}}$ during one rotation period $T_{\rm rot}=2\pi/\omega$ by keeping $\lambda_s^0(t)$ fixed during this rotation and integrating the torque in the body frame $\vec{T}_{\rm bf}$, parametrized by the subsolar longitude and the Fourier components from Eq.~(\ref{eq:FBF})
\begin{eqnarray}\nonumber
\boldsymbol{\Delta} \tilde{\vec{L}}(t)
&=&\int_0^{T_{\rm rot}}{\rm d}t'\,
\left(\!\!\!
\begin{array}{rrr}
  \cos (\omega t') & -\sin (\omega t') & 0 \\
  \sin (\omega t') &  \cos (\omega t') & 0 \\
 0 & 0 & 1 \\
\end{array}
\!\!\!
\right) 
\vec{T}_{\rm bf}(t')\\\nonumber
&=&\frac{T_{\rm rot}}{2\pi}\int_0^{2\pi}\!\!\!{\rm d} \lambda_s  \left(\!\!\!
\begin{array}{rrr}
 \cos (\lambda_s-\lambda_s^0) & \sin (\lambda_s-\lambda_s^0) & 0 \\
 -\sin (\lambda_s-\lambda_s^0) & \cos (\lambda_s-\lambda_s^0) & 0 \\
 0 & 0 & 1 \\
\end{array}
\!\!\!
\right)
\vec{T}_{\rm bf}(t,\lambda_s)\\\nonumber
&=&\frac{T_{\rm rot}}{2}\left(
\begin{array}{c}
-(C_{x,1}-C_{y,2}) \sin \lambda_s^0-(C_{x,2}+C_{y,1}) \cos \lambda_s^0\\
+(C_{x,1}-C_{y,2}) \cos \lambda_s^0-(C_{x,2}+C_{y,1}) \sin \lambda_s^0\\
2 C_{z,0} 
\end{array}
\right)\\
&=&\frac{T_{\rm rot}}{2}
\underbrace{\left(
\begin{array}{c}
C_{x,1}-C_{y,2}\\
C_{x,2}+C_{y,1}\\
2 C_{z,0} 
\end{array}
\right)}_{\text{shape}}
\underbrace{\left(
\begin{array}{rrr}
\sin \lambda_s^0&-\cos \lambda_s^0&0\\
\cos \lambda_s^0& \sin \lambda_s^0&0\\
0 & 0 &1 
\end{array}
\right)}_{\text{orbit}}.
\label{eq:Lapprox}
\end{eqnarray}
The angular momentum change along the entire orbit is then apprximated by adding all $\tens{R}(t_S) \Delta \tilde{\vec{L}}(t)$ contributions to the initial angular momentum.
The ``orbit'' matrix does not affect the magnitude of the ``shape'' vector.
All shown results are done without this approximation and use the full numerical solution of Eq.~(\ref{eq:rotdyn}).
Eq.~(\ref{eq:Lapprox}) is used to determine the physically relevant Fourier components 
\begin{eqnarray}\label{eq:FCrobust}
C_{I}(t)&=&C_{x,1}(t)-C_{y,2}(t)\nonumber\\
C_{II}(t)&=&C_{x,2}(t)+C_{y,1}(t)\\
C_{III}(t)&=&C_{z,0}(t)\nonumber
\end{eqnarray}
for analyzing the observations.

\subsection{Extract observed torque from the rotation-axis evolution}

Next, we consider the inverse problem of finding a plausible torque function in the cometary body frame as function of subsolar coordinates and solar distance.
We infer the torque in the body frame from the observation as function of time $t$ under the assumption of an initial alignment of rotation axis and angular momentum, and with the tensor of inertia given in Eq.~(\ref{eq:inertia}).

To parametrize the observed torque as function of subsolar longitude, we compute  $\lambda_s(t)$ from Eq.~(\ref{eq:lambdas}) at each observation time.
Every 10 days, we find the closest instance $t_i$ of $\lambda_s(t_i)=0$ and compute the Fourier coefficients $C_{x,y,z}^{\rm obs}$ to represent $\vec{T}_{\rm bf}(\lambda_s=0\ldots 2\pi)$.
Only the three Fourier combinations from  Eq.~(\ref{eq:FCrobust}) components should be retrieved from the fit (Fig.~\ref{fig:robustFC}), since the axis motion is not sensitive to the other Fourier components (see Eq.~(\ref{eq:Lapprox})).

\section{Matching Fourier coefficients with the observed torque}

The simplest sublimation model~A results in a rotation axis movement shown in Fig.~\ref{fig:resultPole}, green line.
The evolution is rotated by 90$^\circ$ with respect to the observed torque movement (Fig.~\ref{fig:resultPole}, red line) and leads to a largely increased axis tilt compared to observations.
To explain the observations requires to consider a spatially heterogeneous surface with varying water-ice coverage (model~P).
We could show that an alternative explanation is a large thermal lag of several hours of the maximum sublimation with respect to the maximum irradiation caused by a dust layer of some millimeters thickness.
However, this scenario is unlikely, since the response of sublimation rates to radiation is almost immediate as seen by short ($<1$~h) delays of jet outbreaks (\cite{Lai2016,Shi2016}) and the good agreement of inner dust structures with illumination driven dust release (\cite{Kramer2016,Kramer2018}).
{\color{black}
Measurements of VIRTIS and MIRO found that the thermal inertia is lower than $320$~JK$^{-1}$m$^{-2}$s$^{-0.5}$ when including the error bars (see \cite{Marshall2018} for an overview).
Such a small thermal inertia is not able to provide the needed phase lag of several hours of the maximum sublimation with respect to the maximum irradiation as thermal simulations show and measurements of the activity maxima compared to noon time show \cite{Shi2016}.
}
In fact we show that the often invoked unrealistic thermal lag to explain the non-gravitational forces acting on the cometary orbit can be at least partly replaced by the effects of a complex nucleus shape and its slightly non-uniform activity (\cite{Davidsson2005},\cite{Sosa2009}).

\begin{figure*}
\begin{center}
\includegraphics[width=0.9\linewidth]{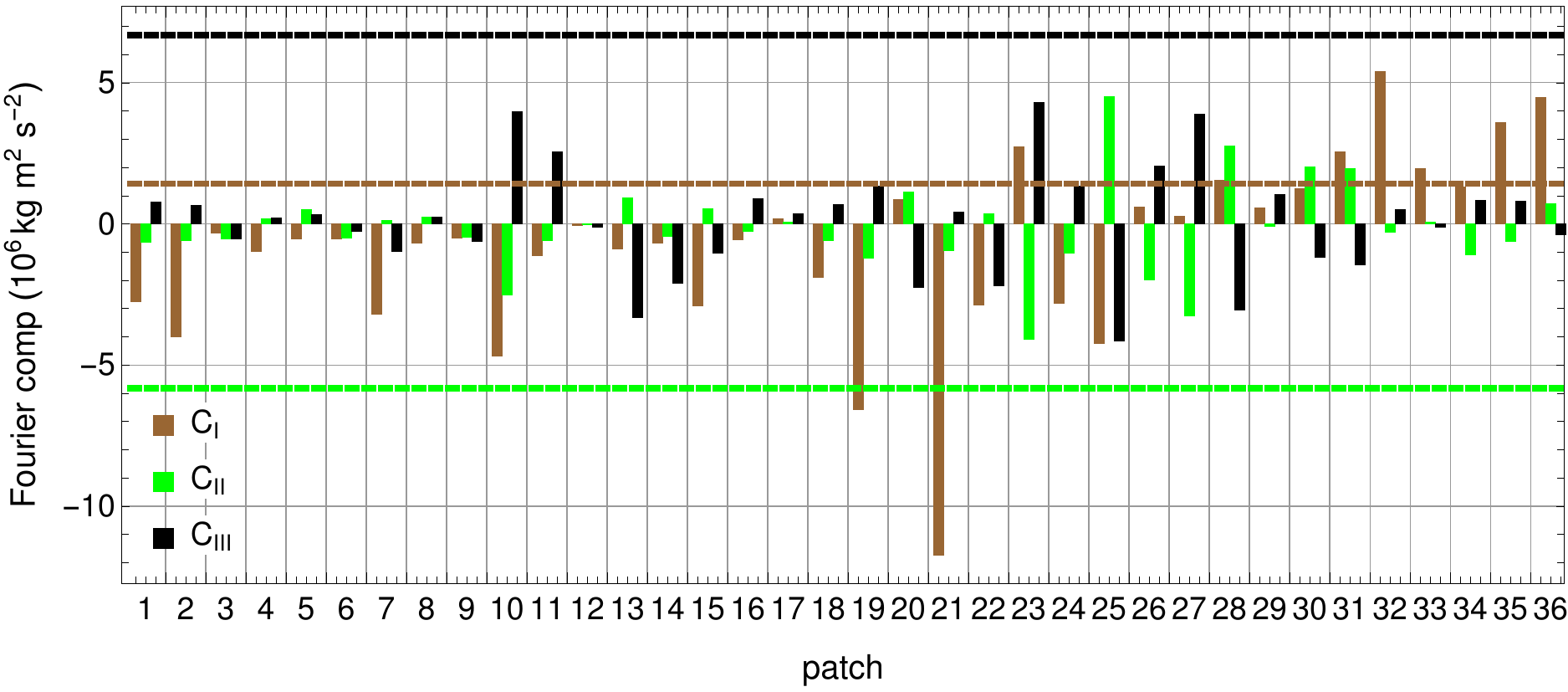}\\
\end{center}
\caption{
{\color{black}
Influence of the different surface areas on the torque evolution for a uniform active surface.
Shown are the extrema of the Fourier torque components for each surface patch (see Fig.~\ref{fig:awf} for the patch boundaries).
The model P seeks linear combinations of patches which in sum match the extrema derived from the observation, indicated by the dashed lines. 
}
}\label{fig:patchescontribution}
\end{figure*}

For the non-uniform case, we divide the surface in 36 equally spaced patches and compute their separate contributions to the torque using the Fourier method described before. 

{\color{black}
Each patch provides a specific contribution to the Fourier components $C_I$, $C_{II}$, $C_{III}$ in Eq.~(\ref{eq:FCrobust}) of the complete comet.
For a uniform activity, the resulting extrema of the Fourier components are shown in Fig.~\ref{fig:patchescontribution} for each patch.
To match the observed rotation state, a linear combination of the patch contributions must yield the observed values of $C_I$, $C_{II}$, and $C_{III}$ in Fig.~\ref{fig:robustFC}, indicated by the dashed lines in Fig.~\ref{fig:patchescontribution}.
The relative ratio of the three components for a single patch is a prescribed property of the sublimation curve. The largest difference of a single patch contribution to the observation is that for the component $C_I$ on patch~21.
The activity of patch~21 has to be reduced, while patches 26-36 with opposite sign for $C_I$ are candidates for an increased activity.
Additional constraints on the activity arise from the simultaneous fitting of the $C_{II}$ and$C_{III}$ components.
To find the activity across all patches we minimize the deviation of observed torque and observations every 20 days with respect to the $L^1$ norm. Details of the data selection and the chosen norm influence the final fit result, but the general structure with the identified depleted and enhanced surface active regions remains unaffected.
}
The fit leads to a closer alignment of observation and model~A/patches for the axis movement (Fig.~\ref{fig:resultPole}), but does not fix the exponent of the total production rate, Fig.~\ref{fig:resultPeriodProduction}, which remains at $Q_{\rm tot}(r)\sim r^{-2.8}$.
In contrast, observations from COPS/DFMS point to a larger exponent $\alpha\sim -6$ to $-7$.
The change of sublimation with heliocentric distance is directly reflected by a small southern excursion of the rotation axis ($300$-$100$~days) before perihelion.
The observations show that the sublimation activity increases non-linearly with insolation, as discussed in Sect.~\ref{subsec:sublimation}.
The effective sublimation curve in Fig.~\ref{fig:subcurve}(a), dashed line yields the total production displayed in Fig.~\ref{fig:resultPeriodProduction}, with larger exponent $\alpha<-5$ as measured by several Rosetta instruments (see e.g.\  \cite{Hansen2016,Kramer2017,Lauter2018}) and modeled by \cite{Hu2017}.
The rotation-axis motion of this modified sublimation model is shown in Fig.~\ref{fig:resultPole} and are in better agreement with observations than the other considered scenarios.

\begin{figure}
\begin{center}
\includegraphics[width=0.7\linewidth]{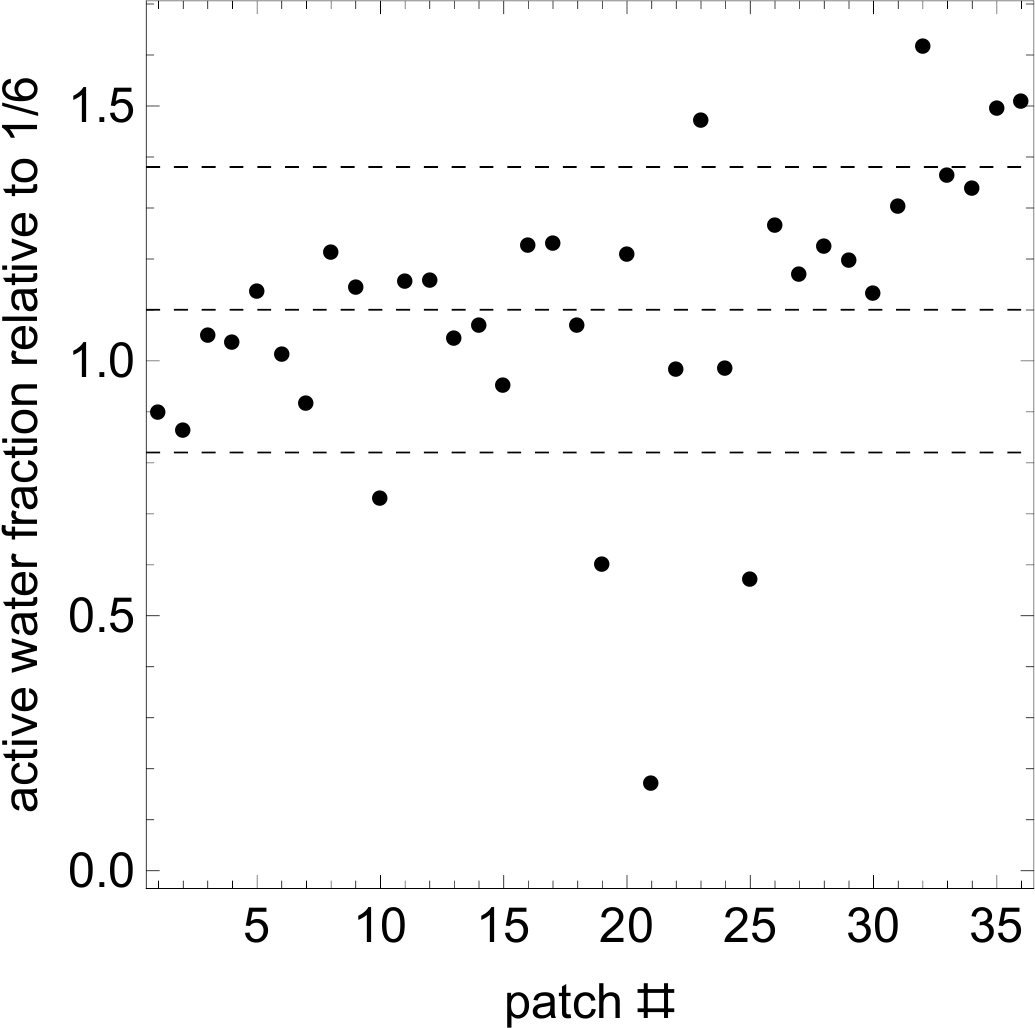}\\
\end{center}
\caption{
Surface active fraction $f_i$ (Eq.~\ref{eq:Tbf}) relative to $1/6$ determined from the torque fit using the effective sublimation curve from Fig.~\ref{fig:subcurve}(a), dashed line, with 36 patches shown in Fig.~\ref{fig:awf}.
The dashed lines indicate the mean value and the standard deviation.
}\label{fig:awfbc}
\end{figure}

\begin{figure*}
\begin{center}
\includegraphics[width=0.7\linewidth]{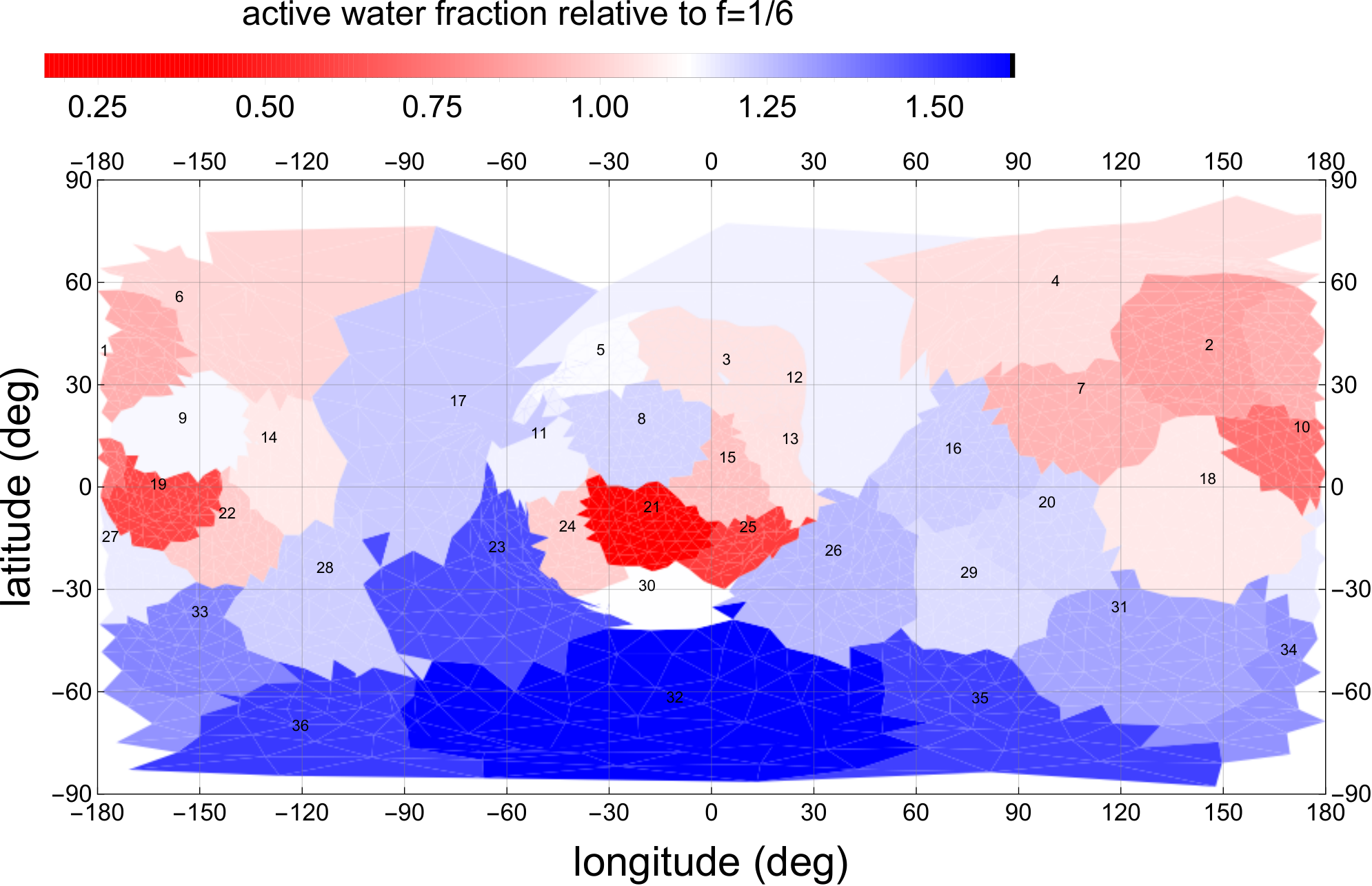}
\end{center}
\caption{
Surface map showing the surface active fraction $f_i$ (Eq.~\ref{eq:Tbf}) relative to $1/6$ corresponding to Fig.~\ref{fig:awfbc}.
The numbers indicate the patch label.
}\label{fig:awf}
\end{figure*}

\begin{figure*}
\begin{center}
\includegraphics[width=0.83\linewidth]{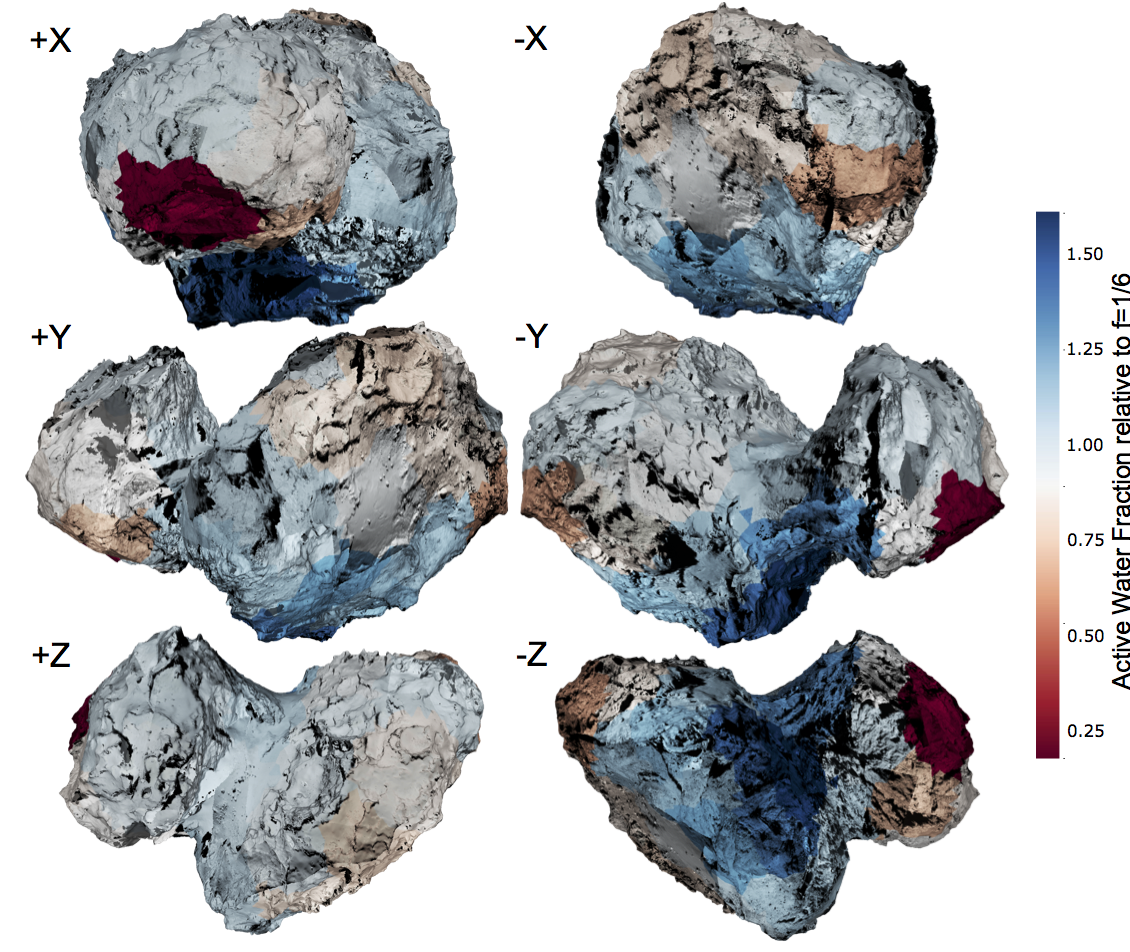}
\end{center}
\caption{The surface active fraction map projected onto the DLR SHAP7 shape model (Preusker et al, 2017).
The shape has been textured using 30 OSIRIS NAC images acquired during the SHAP4S, SHAP5 mission phases for the norther hemisphere and the SHAP7 and SHAP8 mission phases for the southern hemisphere.
The color overlay shows the active surface fraction from Fig.~\ref{fig:awf}
{\color{black}with the view vector indicated by the basis vectors $\vec{X},\vec{Y},\vec{Z}$ in the body frame.}
}
\label{fig:surface1}
\end{figure*}

\section{Implication for the surface composition}

\begin{figure*}
\begin{center}
\includegraphics[width=0.84\linewidth]{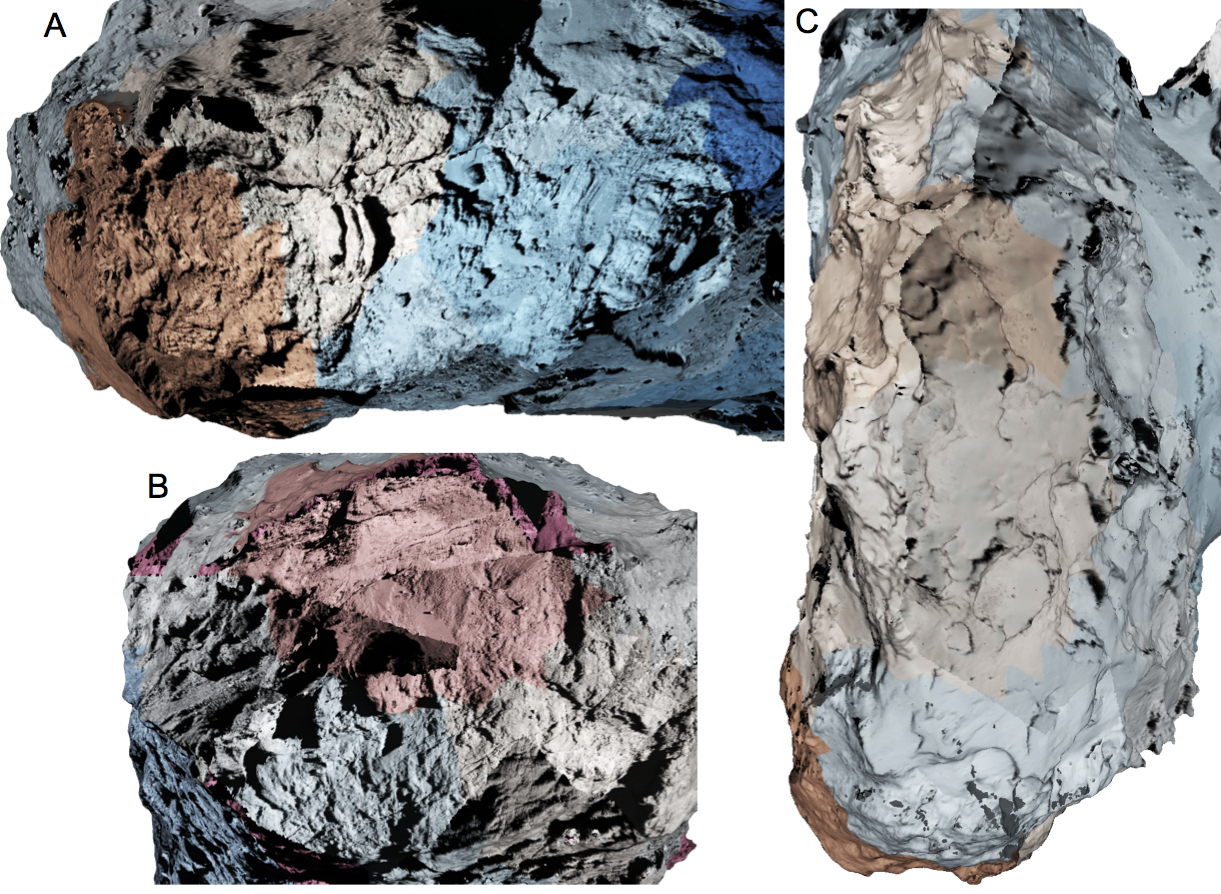}
\end{center}
\caption{Zoom into the shape shown in Fig.~\ref{fig:surface1}.
(A) shows the Khonsu region on the big lobe of the comet nucleus. 
(B) shows the Wosret region of the  small lobe of the comet nucleus and (C) shows the north edge of the big lobe. 
Note that the color intensity of the Wosret (B) view has been decreased as compared to Fig.~\ref{fig:surface1} to allow better visibility of the background image data. The wedge like feature from the left side is an image artifact caused by an image acquired with a high ($\sim 90^\circ$) Sun incidence angle.
}
\label{fig:surface2}
\end{figure*}

{\color{black}The surface active fraction of the best-fit model is shown in Fig.~\ref{fig:awfbc} and as planar map in Fig.~\ref{fig:awf}.
The maps show the active fraction relative to the mean active fraction to highlight the differences to a uniformly active surface.
The absolute value of the surface activity depends on the precise values of the cometary mass and the sublimation curve, while the relative distribution is not strongly affected.
}
{\color{black}
Patches with increased active water fraction are located in the southern hemisphere which agrees with the activity shown in Fig.~6 by \cite{Fougere2016a} derived from Rosetta ROSINA/COPS/DFMS in-situ gas densities.
The direct use measured gas densities from the ROSINA instruments to constrain the diurnal activity and the rotation state is limited, since for operational reasons Rosetta predominantly sampled gas in terminator illumination.
}
Overall, the standard deviation from the homogeneous active surface (mean value $1.10$) is $0.28$, with the smallest activity on patch~21 (six times reduced active surface fraction).
This confirms that all the surface of 67P/C-G shows activity whenever insolated.

A detailed correlation of our 36 patches with all geological regions cannot be expected since the resolution is just not good enough considering that the number of defined regions are now about {\color{black}twice as large} \cite{Thomas2018}.
In Fig.~\ref{fig:surface1} the surface active-fraction regions from Fig.~\ref{fig:awf} have been draped onto the shape model of 67P/C-G (shown as color overlay).
30 OSIRIS NAC images have been mapped onto the shape to provide the morphological context.
The images have been acquired during the SHAP4S and SHAP5 mission phases for the northern hemisphere (September to October 2014), and SHAP7 and SHAP8 for the southern hemisphere (April to June 2016).
Some image boundaries are visible in the mosaic because of the varying illumination condition present during the mission phases.
In general the surface active fraction shows a north south trend with the highest active fraction being in the rough consolidated terrain of the south oriented regions (In particular around the southern neck regions, Fig.~\ref{fig:surface1}~$-Z$). 
The northern dust covered regions like the Seth and Hapi region in the northern neck (Fig.~\ref{fig:surface1}~$+Z$) shows intermediate levels of active fraction.
This is compatible with the northern neck region being the most active in dust production during the early parts of the Rosetta mission. 
Some other features are seen: 
The active fraction map shows a dichotomy between northern neck region of the big lobe (Seth) and the northern foot regions of the big lobe (see Fig.~\ref{fig:surface2}c).
This dichotomy is not reflected in the surface morphology.
Both sides of the big lobe show the same kind of smooth dust covered terrain.
It does, however, make sense from an insolation point of view. The northern neck is in polar night during the perihelion passage while the foot of the big lobe is permanently illuminated throughout the comet year (\cite{Keller2015a}).
The volatiles in the northern neck are being replenished by seasonal mass transport on the comet (\cite{Keller2017}).
Mass transport on the foot of the comet will tend to accumulate in the Imhotep region which is a gravitational low point on the comet.
The Imhotep region (Fig.~\ref{fig:surface1}~$-X$) does indeed show comparable levels of active fraction to those of the northern neck.
The Khonsu region (Fig.~\ref{fig:surface2}a) shows a south-east to north-west gradient in active fraction.
The Khonsu region is a depression with a very rough terrain.
Khonsu may be the result of an earlier fragmentation event that has caused parts of the surface to break off the nucleus.
There is no significant morphological difference between one end of Khonsu and the other and the integrated insolation is comparable.
This may be a modeling artifact caused by the non random choice of patch boundaries. 
The Wosret region (Fig.~\ref{fig:surface2}b) shows a surprising low level of active fraction.
The Wosret region is the major part of the polar circle that receives permanent diurnal illumination during the perihelion passage of the comet.
The region therefore has the highest potential level of activity of any region on the comet (highest integrated insolation).
The morphology of the region is, however, quite different from the other south oriented regions on the comet.
Wosret has a highly smooth but consolidated terrain towards the top of the small lobe and a much rougher consolidated terrain towards the southern neck.
The areas of the active fraction map with lowest values are correlated with the smooth consolidated terrain.
The rougher parts shows significantly higher active fraction.
These active fraction values are more compatible with the levels found in the southern neck which has comparable terrain morphology.
A possible explanation is that the smooth consolidated terrain is simply depleted of volatiles and will therefore exhibit no activity no matter the insolation.
The smooth consolidated Wosret region could represents the final state terrain of cometary evolution.

\section{Conclusions}

We have presented a method to parametrize the observed rotation-axis movement in terms of a theory of Fourier coefficients.
The sublimation induced torques are encoded in three physical relevant combinations of the Fourier coefficients, which steer the rotation period changes and the rotation axis movement.
In particular, the rotation state of 67P/C-G is determined from the orbital evolution of the subsolar longitude and the specific shape.
The increase of the rotation period is caused by the diurnal-average of the rotation axis aligned torque (Fourier coefficient $C_{III}=C_{0,z}$), while the orientation change is caused by the diurnal torque cycle of the perpendicular components (Fourier coefficients $C_{I}$, $C_{II}$).
Only by taking all three Fourier components together, a consistent fit results which constrains the local surface active fraction.
From our analysis we conclude the following points:
\begin{itemize}
    \item The sublimation model~P contains a best-fit for the surface active fraction to the observed 
    rotation state, namely period and axis orientation.
    \item The model includes a sublimation curve that increases much faster than linearly with insolation and reproduces the water production of 67P/C-G in \cite{Hansen2016,Lauter2018}.
    \item A relatively small local variability (standard deviation $0.28$) of the active surface fraction yields the required changes of the rotation state.
    \item Some area around Wosret on the small lobe seems to be less active, while the southern latitudes $<-60^\circ$ show an increased surface active fraction. 
\end{itemize}
A further argument for a mostly uniform gas release comes from the observation of the dust structures in the inner coma modeled by \cite{Kramer2016,Kramer2018}.
The developed Fourier theory could be applied to other solar system bodies, for which accurate measurements of the rotation axis motion and the shape are available.

\begin{acknowledgements}
The authors acknowledge the North-German Supercomputing Alliance (HLRN) for providing computing time on the Cray XC40.
\end{acknowledgements}

%-------------------------------------------------------------------

\bibliographystyle{aa}

\end{document}